\documentstyle[12pt]{article}
\newcommand{\nc}{\newcommand}
\nc{\renc}{\renewcommand}

\nc{\half}{{\textstyle{1\over2}}}
\nc{\etal}{\mbox{\it et al. }}
\nc{\ie}{{\it i.e.}}
\nc{\eg}{{\it e.g.}}

\renc{\thefootnote}{\arabic{footnote}}
\nc{\capt}[1]{{\bf Figure.} {\small\sl #1}}

\nc{\eqs}[2]{\mbox{Eqs.~(\ref{#1},\,\ref{#2})}}
\nc{\eq}[1]{\mbox{Eq.~(\ref{#1})}}

\nc{\figs}[2]{\mbox{Figs.~(\ref{#1},\,\ref{#2})}}
\nc{\fig}[1]{\mbox{Fig~.(\ref{#1})}}

\nc{\tag}[1]{\label{#1} \marginpar{{\footnotesize #1}}}
\nc{\mtag}[1]{\label{#1} \mbox{\marginpar{{\footnotesize #1}}}}
\renc{\baselinestretch}{1.5}
\newlength{\overeqskip}
\newlength{\undereqskip}
\nc{\be}[1]{\begin{equation} \mbox{$\label{#1}$}}
\nc{\bea}[1]{\begin{eqnarray} \mbox{$\label{#1}$}}
\nc{\Section}[2]{\section{#2}\label{#1}}
\nc{\Bibitem}[1]{\bibitem{#1}}
\nc{\Label}[1]{\label{#1}}
\nc{\eea}{\vspace{\undereqskip}\end{eqnarray}}
\nc{\ee}{\vspace{\undereqskip}\end{equation}}
\nc{\bdm}{\begin{displaymath}}
\nc{\edm}{\end{displaymath}}
\nc{\dpsty}{\displaystyle}
\nc{\bc}{\begin{center}}
\nc{\ec}{\end{center}}
\nc{\ba}{\begin{array}}
\nc{\ea}{\end{array}}
\nc{\bab}{\begin{abstract}}
\nc{\eab}{\end{abstract}}
\nc{\btab}{\begin{tabular}}
\nc{\etab}{\end{tabular}}
\nc{\bit}{\begin{itemize}}
\nc{\eit}{\end{itemize}}
\nc{\ben}{\begin{enumerate}}
\nc{\een}{\end{enumerate}}
\nc{\bfig}{\begin{figure}}
\nc{\efig}{\end{figure}}
\nc{\arreq}{&\!=\!&}
\nc{\arrmi}{&\!-\!&}
\nc{\arrpl}{&\!+\!&}
\nc{\arrap}{&\!\!\!\approx\!\!\!&}
\nc{\non}{\nonumber\\*}
\nc{\align}{\!\!\!\!\!\!\!\!&&}

\def\lsim{\; \raise0.3ex\hbox{$<$\kern-0.75em
      \raise-1.1ex\hbox{$\sim$}}\; }
\def\gsim{\; \raise0.3ex\hbox{$>$\kern-0.75em
      \raise-1.1ex\hbox{$\sim$}}\; }
\nc{\DOT}{\hspace{-0.08in}{\bf .}\hspace{0.1in}}
\nc{\Laada}{\hbox {$\sqcap$ \kern -1em $\sqcup$}}
\nc\loota{{\scriptstyle\sqcap\kern-0.55em\hbox{$\scriptstyle\sqcup$}}}
\nc\Loota{{\sqcap\kern-0.65em\hbox{$\sqcup$}}}
\nc\laada{\Loota}
\nc{\qed}{\hskip 3em \hbox{\BOX} \vskip 2ex}

\nc{\real}{{\rm I \! R}}
\nc{\Z}{{\sf Z \!\!\! Z}}
\nc{\complex}{{\rm C\!\!\! {\sf I}\,\,}}
\def\bigid{\leavevmode\hbox{\small1\kern-3.8pt\normalsize1}}
\def\id{\leavevmode\hbox{\small1\kern-3.3pt\normalsize1}}
\nc{\slask}{\!\!\!/}
\nc{\bis}{{\prime\prime}}
\nc{\pa}{\partial}
\nc{\na}{\nabla}
\nc{\ra}{\rangle}
\nc{\la}{\langle}
\nc{\goto}{\rightarrow}
\nc{\swap}{\leftrightarrow}

\nc{\EE}[1]{ \mbox{$\cdot10^{#1}$} }
\nc{\abs}[1]{\left|#1\right|}
\nc{\at}[2]{\left.#1\right|_{#2}}
\nc{\norm}[1]{\|#1\|}
\nc{\abscut}[2]{\Abs{#1}_{\scriptscriptstyle#2}}
\nc{\vek}[1]{{\rm\bf #1}}
\nc{\integral}[2]{\int\limits_{#1}^{#2}}
\nc{\inv}[1]{\frac{1}{#1}}
\nc{\dd}[2]{{{\partial #1}\over{\partial #2}}}
\nc{\ddd}[2]{{{{\partial}^2 #1}\over{\partial {#2}^2}}}
\nc{\dddd}[3]{{{{\partial}^2 #1}\over
	{\partial #2 \partial #3}}}
\nc{\dder}[2]{{{d #1}\over{d #2}}}
\nc{\ddder}[2]{{{d^2 #1}\over{d {#2}^2}}}
\nc{\dddder}[3]{{d^2 #1}\over
	{d #2 d #3}}
\nc{\dx}[1]{d\,^{#1}x}
\nc{\dy}[1]{d\,^{#1}y}
\nc{\dz}[1]{d\,^{#1}z}
\nc{\dl}[1]{\frac{d\,^{#1}l}{(2\pi)^{#1}}}
\nc{\dk}[1]{\frac{d\,^{#1}k}{(2\pi)^{#1}}}
\nc{\dq}[1]{\frac{d\,^{#1}q}{(2\pi)^{#1}}}

\nc{\cc}{\mbox{$c.c.$ }}
\nc{\hc}{\mbox{$h.c.$ }}
\nc{\cf}{cf.\ }
\nc{\erfc}{{\rm erfc}}
\nc{\Tr}{{\rm Tr\,}}
\nc{\tr}{{\rm tr\,}}
\nc{\pol}{{\rm pol}}
\nc{\sign}{{\rm sign}}
\nc{\bfT}{{\bf T }}

\def\GeV{{\rm\ GeV}}

\nc{\cA}{{\cal A}}
\nc{\cB}{{\cal B}}
\nc{\cD}{{\cal D}}
\nc{\cE}{{\cal E}}
\nc{\cG}{{\cal G}}
\nc{\cH}{{\cal H}}
\nc{\cL}{{\cal L}}
\nc{\cO}{{\cal O}}
\nc{\cT}{{\cal T}}
\nc{\cN}{{\cal N}}
\nc{\rvac}[1]{|{\cal O}#1\rangle}
\nc{\lvac}[1]{\langle{\cal O}#1|}
\nc{\rvacb}[1]{|{\cal O}_\beta #1\rangle}
\nc{\lvacb}[1]{\langle{\cal O}_\beta #1 |}
\nc{\bb}{\bar{\beta}}
\nc{\bt}{\tilde{\beta}}
\nc{\ctH}{\tilde{\cal H}}
\nc{\chH}{\hat{\cal H}}
\nc{\1}{\aa}
\nc{\2}{\"{a}}
\nc{\3}{\"{o}}
\nc{\4}{\AA}
\nc{\5}{\"{A}}
\nc{\6}{\"{O}}
\nc{\al}{\alpha}
\nc{\g}{\gamma}
\nc{\Del}{\Delta}
\nc{\e}{\epsilon}
\nc{\eps}{\epsilon}
\nc{\lam}{\lambda}
\nc{\om}{\omega}
\nc{\Om}{\Omega}
\nc{\ve}{\varepsilon}
\nc{\mn}{{\mu\nu}}
\nc{\k}{\kappa}
\nc{\vp}{\varphi}

\nc{\advp}[3]{{\it  Adv.\ in\ Phys.\ }{{\bf #1} {(#2)} {#3}}}
\nc{\annp}[3]{{\it  Ann.\ Phys.\ (N.Y.)\ }{{\bf #1} {(#2)} {#3}}}
\nc{\apl}[3]{{\it  Appl. Phys. Lett. }{{\bf #1} {(#2)} {#3}}}
\nc{\apj}[3]{{\it  Ap.\ J.\ }{{\bf #1} {(#2)} {#3}}}
\nc{\apjl}[3]{{\it  Ap.\ J.\ Lett.\ }{{\bf #1} {(#2)} {#3}}}
\nc{\app}[3]{{\it Astropart.\ Phys.\ }{{\bf #1} {(#2)} {#3}}}
\nc{\cmp}[3]{{\it  Comm.\ Math.\ Phys.\ }{{ \bf #1} {(#2)} {#3}}}
\nc{\cqg}[3]{{\it  Class.\ Quant.\ Grav.\ }{{\bf #1} {(#2)} {#3}}}
\nc{\epl}[3]{{\it  Europhys.\ Lett.\ }{{\bf #1} {(#2)} {#3}}}
\nc{\ijmp}[3]{{\it Int.\ J.\ Mod.\ Phys.\ }{{\bf #1} {(#2)} {#3}}}
\nc{\ijtp}[3]{{\it Int.\ J.\ Theor.\ Phys.\ }{{\bf #1} {(#2)} {#3}}}
\nc{\jmp}[3]{{\it  J.\ Math.\ Phys.\ }{{ \bf #1} {(#2)} {#3}}}
\nc{\jpa}[3]{{\it  J.\ Phys.\ A\ }{{\bf #1} {(#2)} {#3}}}
\nc{\jpc}[3]{{\it  J.\ Phys.\ C\ }{{\bf #1} {(#2)} {#3}}}
\nc{\jap}[3]{{\it J.\ Appl.\ Phys.\ }{{\bf #1} {(#2)} {#3}}}
\nc{\jpsj}[3]{{\it J.\ Phys.\ Soc.\ Japan\ }{{\bf #1} {(#2)} {#3}}}
\nc{\lmp}[3]{{\it Lett.\ Math.\ Phys.\ }{{\bf #1} {(#2)} {#3}}}
\nc{\mpl}[3]{{\it  Mod.\ Phys.\ Lett.\ }{{\bf #1} {(#2)} {#3}}}
\nc{\ncim}[3]{{\it  Nuov.\ Cim.\ }{{\bf #1} {(#2)} {#3}}}
\nc{\np}[3]{{\it  Nucl.\ Phys.\ }{{\bf #1} {(#2)} {#3}}}
\nc{\npps}[3]{{\it  Nucl.\ Phys.\ Proc.\ Suppl.\ }{{\bf #1} {(#2)} {#3}}}
\nc{\pr}[3]{{\it Phys.\ Rev.\ }{{\bf #1} {(#2)} {#3}}}
\nc{\pra}[3]{{\it  Phys.\ Rev.\ A\ }{{\bf #1} {(#2)} {#3}}}
\nc{\prb}[3]{{\it  Phys.\ Rev.\ B\ }{{{\bf #1} {(#2)} {#3}}}}
\nc{\prc}[3]{{\it  Phys.\ Rev.\ C\ }{{\bf #1} {(#2)} {#3}}}
\nc{\prd}[3]{{\it  Phys.\ Rev.\ D\ }{{\bf #1} {(#2)} {#3}}}
\nc{\prl}[3]{{\it Phys.\ Rev.\ Lett.\ }{{\bf #1} {(#2)} {#3}}}
\nc{\pl}[3]{{\it  Phys.\ Lett.\ }{{\bf #1} {(#2)} {#3}}}
\nc{\prep}[3]{{\it Phys.\ Rep.\ }{{\bf #1} {(#2)} {#3}}}
\nc{\prsl}[3]{{\it Proc.\ R.\ Soc.\ London\ }{{\bf #1} {(#2)} {#3}}}
\nc{\ptp}[3]{{\it  Prog.\ Theor.\ Phys.\ }{{\bf #1} {(#2)} {#3}}}
\nc{\ptps}[3]{{\it  Prog\ Theor.\ Phys.\ suppl.\ }{{\bf #1} {(#2)} {#3}}}
\nc{\physa}[3]{{\it  Physica\ A\ }{{\bf #1} {(#2)} {#3}}}
\nc{\physb}[3]{{\it  Physica\ B\ }{{\bf #1} {(#2)} {#3}}}
\nc{\phys}[3]{{\it Physica\ }{{\bf #1} {(#2)} {#3}}}
\nc{\rmp}[3]{{\it  Rev.\ Mod.\ Phys.\ }{{\bf #1} {(#2)} {#3}}}
\nc{\rpp}[3]{{\it Rep.\ Prog.\ Phys.\ }{{\bf #1} {(#2)} {#3}}}
\nc{\sjnp}[3]{{\it Sov.\ J.\ Nucl.\ Phys.\ }{{\bf #1} {(#2)} {#3}}}
\nc{\spjetp}[3]{{\it Sov.\ Phys.\ JETP\ }{{\bf #1} {(#2)} {#3}}}
\nc{\yf}[3]{{\it Yad.\ Fiz.\ }{{\bf #1} {(#2)} {#3}}}
\nc{\zetp}[3]{{\it Zh.\ Eksp.\ Teor.\ Fiz.\  }{{\bf #1}  {(#2)} {#3}}}
\nc{\zp}[3]{{\it Z.\ Phys.\ }{{\bf #1} {(#2)} {#3}}}
\nc{\ibid}[3]{{\sl ibid.\ }{{\bf #1} {#2} {#3}}}
\nc{\rf}[1]{(\ref{#1})}
\nc{\nn}{\nonumber \\*}
\nc{\bfB}{\bf{B}}
\nc{\bfv}{\bf{v}}
\nc{\bfx}{\bf{x}}
\nc{\bfy}{\bf{y}}
\nc{\vx}{\vec{x}}
\nc{\vy}{\vec{y}}
\nc{\oB}{\overline{B}}
\nc{\oI}{\overline{I}}
\nc{\oR}{\overline{R}}
\nc{\rar}{\rightarrow}
\nc{\ti}{\times}
\nc{\slsh}{\hskip-5pt/}
\nc{\sm}{Standard~Model~}
\nc{\MP}{M_{\rm Pl}}
\nc{\tp}{t_{\rm Pl}}
\nc{\ave}{\bar{E}}

\nc{\eff}{{\rm eff}}
\nc{\kk}{\vek{k}}
\nc{\pp}{{\rm p}}
\nc{\ga}{g_{a\gamma}}
\nc{\vv}{\\}
\nc{\eee}{{\bf E}}
\nc{\bbb}{{\bf B}}
\nc{\qcd}{T_{\rm QCD}}
\nc{\G}{\rm \ G}
\def\vec#1{{\bf #1}}
\def\lae{\;^{<}_{\sim} \;} \def\gae{\; ^{>}_{\sim} \;} 

\def\udd{u^{c}d^{c}d^{c}}

\def\uude{u^{c}u^{c}d^{c}e^{c}}

\begin{document}
{\title{\vskip-2truecm{\hfill {{\small \\
	\hfill HIP-1998-30/th \\
	}}\vskip 1truecm}
{\bf  Supersymmetric D-term Inflation and B-ball Baryogenesis}}
%\vspace{1.2cm}
{\author{
{\sc  Kari Enqvist$^{1}$}\\
{\sl\small Department of Physics and Helsinki Institute of Physics,}\\ 
{\sl\small P.O. Box 9,
FIN-00014 University of Helsinki,
Finland}\\
{\sc and}\\
{\sc  John McDonald$^{2}$}\\
{\sl\small Department of Physics, P.O. Box 9,
FIN-00014 University of Helsinki,
Finland}
}
\maketitle
%\vspace{1cm}
%\newpage
\begin{abstract}
\noindent
We consider the B-ball cosmology of the MSSM in the context of D-term inflation models where the reheating temperature 
is determined by the Affleck-Dine mechanism to be of the order of 1 GeV. We show that such a low reheating temperature 
can arise quite naturally as the result of a symmetry which is required to
 maintain the flatness of the inflaton potential. In this case the B-balls will decay well after the freeze-out 
temperature of the LSP, allowing baryons and cold dark matter to originate primarily from B-ball decays.
\end{abstract}
\vfil
\footnoterule
{\small $^1$enqvist@pcu.helsinki.fi};
{\small $^2$mcdonald@rock.helsinki.fi}

\thispagestyle{empty}
\newpage
\setcounter{page}{1}

%%%%%%%%%%%%%%%%%%%%%%%%%%%%%%%%%%%%%%%%%%%%
%%%%%%%%%%%%%%%%%%%%%%%%%%%%%%%%%%
%%%%%%%%%%%%%%%%%%%%%%%%%%%%%%%%%%%%%%%%%%%
%%%%%%%%%%%%%%%%%%%%%%%%%%%%%%%%%%%
The MSSM \cite{nilles} is known to admit non-topological solitons in its spectrum \cite{kuz};  
in particular, B-balls (Q-balls \cite{cole,cole2} with baryonic charge) can exist. 
In a cosmological scenario which includes
inflation these can be copiously produced by the breakdown of scalar 
condensates along flat directions of the MSSM. 
In the case of gauge-mediated SUSY breaking, extensively discussed 
in references [5-10], the B-balls are stable and could account for cold 
dark matter \cite{ks2}. 
In the case of gravity-mediated
breaking, studied in \cite{bbb1,bbb2}, 
the B-balls are unstable\footnote{\small Depending on the details of the hidden sector and its interaction with the observable sector, it is conceivable that SUSY breaking could be 
suppressed for large values of the condensate field,
leading to $stable$ B-balls in the case of gravity-mediated
SUSY breaking \cite{kpriv}.}
but nevertheless, if they can survive thermalization,
they are typically
long-lived enough to decay much after the electroweak phase transition, 
leading to a variant of the Affleck-Dine mechanism \cite{ad} which we call B-ball Baryogenesis (BBB) \cite{bbb1,bbb2}. 
The requirement that they can survive thermalization implies that 
the B-balls in R-parity conserving models originate from a d=6 AD condensate and imposes
 an upper bound on the reheating temperature of $10^{3-5}\GeV$ \cite{bbb1,bbb2}. 
Such B-balls can protect a B asymmetry originating in the AD 
condensate from the effects 
of additional B-L violating interactions, 
which would otherwise wash out the B asymmetry when combined with
anomalous B+L violation \cite{bbb1}. In addition, if the reheating
temperature is sufficiently low that the B-balls decay below the freeze-out temperature of the lightest SUSY particle (LSP), then cold dark matter can mostly come from B-ball decays
rather than from thermal relics. This
opens up the possibility of relating the number density of dark matter particles to that of the baryons, allowing for an explaination of their observed similarity for the case of dark matter particles with weak scale masses \cite{bbb2}. Since BBB requires
 low reheating temperatures, it is important to consider whether
low reheating temperatures are a natural feature of realistic SUSY inflation models. 
We will show that in D-term inflation models \cite{dti,kmr} very low reheating temperatures are indeed a natural feature. 

    In general, SUSY inflation models can be classified as either F- or D-term inflation models, depending on whether the energy density during inflation comes from the F- or D-term contribution to the scalar potential. However, in F-term inflation models, $H$ dependent corrections to the mass terms tend to
spoil the flatness of the inflaton potential \cite{h2o,h2}. As a result, 
the most favoured candidate for SUSY inflation at present is
D-term inflation, which has the form of a
 hybrid inflation scenario \cite{kn,hi}. 
An important feature of D-term inflation models is that 
 the A-terms receive no $H$ dependent corrections \cite{dti,kmr} (although the mass squared terms do receive
 corrections once inflation 
ends \cite{kmr}). This has the important consequence
 that the baryon asymmetry from the d=6 AD condensate will be $maximal$ in a typical domain of the
 Universe, which in turn requires that the reheating temperature is of the order of $1 \GeV$. The main purpose of this paper
 is to consider whether such low reheating temperatures can be a natural feature of realistic D-term inflation
 models, thus making them compatible with BBB and the AD mechanism along d=6 directions.

      The simplest D-term inflation model \cite{dti,kmr} is based on three 
fields: a gauge singlet $S$, which plays the role of the slow-rolling inflaton, and two fields with opposite $U(1)_{FI}$ charges, 
$\psi_{+}$ and $\psi_{-}$, where $U(1)_{FI}$ is the Fayet-Iliopoulos
(FI) gauge group. The superpotential is given by
\be{d1} W = \lambda S \psi_{+} \psi_{-} ~.\ee
The scalar potential, including the $U(1)_{FI}$ D-term, is then
\be{d2} V = |\lambda|^{2}  \left( |
\psi_{+} \psi_{-} |^{2} + |S \psi_{+}|^{2} + |S \psi_{-}|^{2}
\right) + \frac{g^{2}}{2} 
\left( |\psi_{+}|^{2} - |\psi_{-}|^{2} + \xi^{2}\right)^{2}   ~,\ee
where $\xi^{2}$ is the FI term and $g$ is the $U(1)_{FI}$ coupling. $\xi$ is fixed by 
cosmic microwave background radiation (CMBR) measurements \cite{cmbr} to be 
$6.6 \times 10^{15}\GeV$ \cite{drot}. The global minimum of the potential is at 
$S = 0$, $\psi_{+} = 0$, $\psi_{-}= \xi$. However, for $S > S_{crit} \equiv g \xi/\lambda$, the minimum is at $\psi_{+} = \psi_{-} = 0$ and there is a non-zero energy density $\rho = g^{2} \xi^{4}/2$. Although there are no tree-level corrections to the flatness of the $S$ potential due to SUSY breaking by the energy density during inflation, there will be an $S$ potential coming from 1-loop corrections and from higher order terms in $M_{*}^{-1}$ in the K\"ahler potential, 
where $M_{*} = M_{Pl}/\sqrt{8 \pi}$ is the supergravity mass scale \cite{nilles}. The 1-loop term, due to the splitting in mass of the bosonic components of $\psi_{\pm}$ from their fermionic partners, is given by $\Delta V = 
\alpha^{2} \xi^{4} ln(|\lambda S|/\mu)$ \cite{dti,kmr}. 

     In this model, slow-rolling of the $S$ scalar fails well before $S_{crit}$, at $S_{f} \approx \sqrt{\alpha M_{*}^{2}/2 \pi}$.
In order to account for the observed homogeneity and isotropy of the Universe, the slow-roll of $S$ must begin at least at $S_{55}$, where 
\be{d3} S_{55} = \left( \frac{55 \alpha M_{*}}{\pi} + S_{f}^{2} \right)^{1/2} \approx 0.9 M_{*}   ~.\ee      
For such large values of $S$, it is likely that higher order superpotential terms will cause the $S$ scalar potential to deviate from flatness. For example, a term of the form $\Delta W = \kappa S^{m}/M_{*}^{m-3} $ with $S \approx M_{*}$ produces too large a deviation from the observed CMBR unless $\kappa \lae 10^{-5}$. Since we would expect the natural mass scale of the physical interactions coming from this superpotential term to be $M_{*}$, the natural value of $\kappa$ will be of the order of $1/m!$. Thus in order to have a naturally flat potential during inflation, all superpotential terms with $m \leq 9$ must be suppressed by discrete symmetries or R-symmetries \cite{kmr}. R-symmetries are particularly effective in doing this, since they can $completely$ eliminate
higher order superpotential terms which are purely a function of $S$ as well dangerous gauge kinetic terms of the form  $S^{k}W^{\alpha}W_{\alpha}$ \cite{kmr,dlyth}. (There is a danger that a global symmetry could be 
broken by non-perturbative gravitational effects \cite{worm0,worm}, in which case a discrete gauge symmetry would be preferred. However, such effects
 are very sensitive to the details of the gravity theory on small scales, and do not generally exclude models with global symmetries \cite{kallosh}). Depending on the charge of $S$ and the MSSM fields with respect to the R-symmetry or discrete symmetry, this suppression of higher order operators may also suppress the couplings of $S$ to the MSSM sector. This, as we will show, can lead to very low reheating temperatures after the $S$ oscillations decay. 

          In D-term inflation models there are two seperate periods of reheating, corresponding to the decay of the $\psi_{-}$ field (at temperature $T_{R}^\psi$)
and of the inflaton $S$ \cite{kmr}. The initial values of the scalar field expectation values
 at the end of inflation are given by 
$\psi^{'}_{-} \equiv (\xi - \psi_{-}) = \xi$ and $S \approx S_{crit} \equiv g \xi/\lambda$. The masses of the 
scalars $\psi_{-}^{'}$, $\psi_{+}$ and $S$ are $M_{\psi_{-}^{'}} = g \xi$ and 
$M_{S} = M_{\psi_{+}} = \lambda \xi$. Thus the initial energy density in the oscillating 
$\psi_{-}^{'}$ and $S$ fields will be essentially the same, $\rho_{S} \approx \rho_{\psi_{-}^{'}}
\approx g^{2} \xi^{4}$. This has the consequence that the Universe will be matter
 dominated by scalar field oscillations until $both$ condensates have decayed. 
The immediate $\psi_{-}^{'}$ decay leaves the Universe at a temperature $T_R^\psi \approx g^{1/2}\xi$.
 The Universe is subsequently matter dominated, but with an initial radiation density much larger than that expected
 purely from the decays of the inflaton $S$ \footnote{\small The $S$ condensate could be thermalized by this radiation density, via scattering with light MSSM scalars once the $U(1)_{FI}$ gauge fields are integrated out \cite{kmr}. However, thanks to the strong dependence of the scattering rate on $\lambda$ ($\Gamma_{scatt} \propto \lambda^{4}$), this will not occur so long as $\lambda \lae 0.5$.}. 
Eventually, at a temperature $T_{*}$, the radiation coming from the
 decay of the $S$ inflatons, which has an energy density given by
\be{d4} 
\rho_{r} \approx \frac{2 \Gamma_{S} \rho_{S}}{5 H}        
 ~,\ee
 where $\Gamma_{S}$ is the $S$ decay rate,
 comes to dominate that remaining from the first stage of
 reheating, so that the
 subsequent evolution is the same as for the
 minimal single inflaton model of Refs.\cite{bbb1,bbb2}.
The expansion rate at $T_{*}$, $H_{*}$, is given by 
\be{d5}  H_{*} \approx  2 \times 10^{-3} \left( \frac{T_R^\psi}{10^{16}\GeV} \right)^{4/5}
\left( \frac{T_R^S}{10^{2}\GeV}\right)^{6/5}  \GeV    ~.\ee
B-balls typically begin to form at $H_{i} \approx 5 |K|\GeV$, where $|K|$ is the coefficient of the 
logarithmic radiative correction to the B-ball scalar potential; typically $|K| \approx 0.01-0.1$ \cite{bbb1,bbb2}. Thus $H_{*}$
is typically smaller than $H_{i}$, especially for the case of small $T_R^S$.
 This means that the temperature of the Universe when the AD condensate forms (at $H 
\approx 100\GeV$ \cite{bbb1,bbb2}) and when the 
B-balls begin to form at $H_{i}$ will be somewhat higher than in the single inflaton model. However, there is no danger of the B-balls being thermalized 
so long as the AD condensate from which they form is not itself thermalized. This is because the background of non-relativistic squarks and anti-squarks which exists after the AD condensate collapses to B-balls will give large effective masses ($\gg T$) to the particles which could otherwise 
thermalize the B-balls \cite{bbb2}. Therefore the formation and evolution of the 
B-balls in the D-term inflation scenario will be effectively the same as in the minimal single inflaton model. 

                    We next consider the decay of the $S$ scalar oscillations. As noted above, 
the self-couplings of $S$ in the superpotential must be highly suppressed by a symmetry in order to ensure that the 
scalar potential is flat enough for successful inflation. This in turn can suppress the couplings of $S$ to 
MSSM fields. Suppose that the lowest dimension 
 superpotential coupling of the $S$ scalars to the 
MSSM fields is of the form 
\be{d5} W = \frac{\kappa S \phi^{r}}{M_{*}^{r-2}}    ~,\ee
where $\phi$ represents the MSSM fields. This gives, for example, a decay rate to $r$ $\phi$ particles of the form 
\be{d6} \Gamma_{S} \approx \left(\frac{M_{S}}{M_{*}}\right)^{2(r-2)}
\kappa^{2} \beta_{r} M_{S}     ~,\ee
where $\beta_{r}$ represents the phase space factor. The decay temperature is then 
\be{d6}  T_R^S \approx \left(\frac{\sqrt{8 \pi}}{k_{T}} \right)^{1/2}
\left(\frac{M_{S}}{M_{*}} \right)^{r - 5/2}
\kappa \beta_{r}^{1/2} M_{S}    ~,\ee
where during radiation domination $H = k_{T}T^{2}/M_{Pl}$ with $k_{T} \approx 17$.
With $M_{S} = \lambda \xi$ this gives 
\be{d7} T_R^{S} \approx 4 \times 10^{15} (2.8 \times 10^{-3})^{r-5/2} \kappa 
\lambda^{r-3/2} \beta_{r}^{1/2}    ~.\ee
Low reheating temperatures can arise as a result of a "mismatch" between the transformation 
properties of the fields of the inflaton sector ($S$, $\psi_{+}$ and $\psi_{-}$)
 and the MSSM sector under the symmetry.
As a particularly simple example, consider an R-symmetry under which the inflaton sector fields
have charges which are multiples of 1/2 and the MSSM sector have charges which are multiples of 1/3.
An example which allows the $\lambda S \psi_{+} \psi_{-}$ 
superpotential term (allowed terms have a total R-charge equal to 2 \cite{nilles}) 
but eliminates the higher order $S$ superpotential terms is defined by
$R(S) = -n$, $R(\psi_{+}) = R(\psi_{-}) = (n+2)/2$ for the inflaton 
sector fields and
$R(Q) = R(L) = 5/3$, $R(u^{c}) = R(d^{c}) = R(e^{c}) = 1/3$ and 
$R(H_{u}) = R(H_{d}) = 0$ for the MSSM fields. (The MSSM
charges have been chosen in order to allow the
higher order terms which stabilize the d=6 $\udd$ D-flat
direction).
In Table 1 we give the lowest order R-invariant
operators coupling $S$ to the MSSM fields for
the case where $n$ is an positive integer with
$n \leq 3$. (For $n > 3$ the operators are of higher order).
\newpage
\begin{center}{\bf Table 1. R-invariant operators.}\end{center}  

\begin{center}
\begin{tabular}{|c|c|c|}          \hline
$n$ & $r$ & $\phi^{r}$ \\ \hline
$1$ & $6$ &  $(LH_{u})(\uude)$ 
 \\
$2$ & $4$ &  $QQu^{c}d^{c}$ 
 \\
$3$ & $6$ &  $(LH_{u})^{3}$ 
 \\
 \hline
\end{tabular}
\end{center} 
For example, with $n = 1$
the lowest order superpotential coupling of $S$ to the MSSM fields has $r = 6$. 
This gives a reheating temperature  
\be{d8} T_R^S \approx 150 \left(\frac{\lambda}{0.1}\right)^{9/2}
\kappa \beta_{6}^{1/2} \GeV
   ~, \ee
where $\beta_{6} \approx 10^{-6}$.
Thus a low reheating temperature, 
as low as 1 GeV or less, is a natural feature of this model. 

                                       In general, we should also consider 
corrections to the K\"ahler potential and the gauge kinetic term which define the full supergravity theory \cite{nilles}. 
It is straightforward to show that these non-minimal corrections  
do not introduce any operators linear in $S$ of lower order in $M_{*}^{-1}$. It is also possible that such non-minimal corrections could allow 
for rapid decay via parametric resonance \cite{param,param2}.  However, it is straightforward to show \cite{param2} that the resulting resonant decay of the inflaton will be completely negligible so long as $\lambda/g^{1/2}\gae 0.03$. For values of $\lambda$ and $g$ not too small compared with 1 this condition will be easily satisfied.

              We next consider the need for low reheating temperatures in d=6 AD baryogenesis.
 The MSSM has many D-flat directions in the scalar potential, along which the flatness is lifted only by SUSY breaking terms and non-renormalizible superpotential terms \cite{drt}. 
The potential of the complex AD scalar $\Phi$ will have the form
\be{e1} U(\Phi) \approx (m^{2} - c H^{2})\left(1 +  K \log\left( \frac{|\Phi|^{2}}{M^{2}} \right) \right) |\Phi|^{2} 
+ \frac{\lambda^{2}|\Phi|^{2(d-1)}
}{M_{*}^{2(d-3)}} + \left( \frac{A_{\lambda} 
\lambda \Phi^{d}}{d M_{*}^{d-3}} + h.c.\right)    ~,\ee
where $d$ is the dimension of the non-renormalizible term in the superpotential and $c H^{2}$ gives the $H$ correction to the scalar mass (with $c$ positive and of the order of 1).
 The logarithmic correction to the scalar mass term, 
which is expected in SUSY models, is crucial for the growth of perturbations of the AD field
 and the formation of B-balls, which occurs if $K < 0$. Of particular interest to us here will be the $\udd$ direction, for which $K < 0$ and along which a d=6 superpotential term of the form $(\udd)^{2}$ is allowed by the R-symmetry of the previously discussed model. 
This superpotential term (via the associated A-term in the scalar potential) is responsible for
 introducing the B- and CP-violation necessary to produce 
the baryon asymmetry via the AD mechanism \cite{ad}.  
At $c H^{2} > m^{2}$, the 
scalar field will be at the minimum of its potential, with an expectation value given by 
$|\Phi| \equiv \Phi_{o} \approx (c H^{2} M_{*}^{6}/5 \lambda^{2})^{1/8}$. However, in the absence of 
order $H$ corrections to the A-term, the phase of $\Phi$ will be completely undetermined. 
Once $c H^{2} \lae m^{2}$, the AD field will start to oscillate. The A-term then creates a phase shift $\delta$ of the order of 1 between the real and imaginary parts of the oscillating AD field. 
 With initially $\Phi = \Phi_{o} e^{i \delta_{CP}}$, where $\delta_{CP}$ is the random initial phase determined during inflation, the initial B asymmetry in the AD condensate at $H_{o} \approx m/c^{1/2}$ is given by $n_{B\;o} =  \epsilon m \Phi_{o}^{2}/3$, where 
$\epsilon = Sin2\delta_{CP}Sin\delta$. This gives for the baryon to entropy ratio of the d=6 
AD condensate,
\be{e3} \eta_{B} \equiv 
\frac{n_{B}}{s} = \frac{2 \pi n_{B\;o} T_R^S}{H_{o}^{2}M_{p}^{2}} 
 \approx 0.03 c 
\left(\frac{0.008}{\lambda}\right)^{1/2}
\left(\frac{100\GeV}{m}\right)^{1/2}
\left(\frac{T_R^S}{10^{9}\GeV}\right)
~,\ee
where we have assumed that $\lambda \approx 1/5!$, such that the strength of the 
physical self-interactions of $\Phi$ is determined by the mass scale $M_{*}$, and
we have taken $\delta \approx \delta_{CP} \approx 1$. 
The corresponding reheating temperature for the d=6 AD condensate is
\be{e4} T_R^S \approx \frac{3.3}{c}
\left(\frac{\lambda}{0.008}\right)^{1/2}
\left(\frac{m}{100\GeV}\right)^{1/2}
\left(\frac{\eta_{B}}{10^{-10}}\right)
\GeV    ~.\ee 
Nucleosynthesis \cite{sarkar} implies that $\eta_{B} \approx (3-8) \times 10^{-11}$ for an $\Omega = 1$ Universe. Thus, for $\delta_{CP} \approx 1$, we find that $T_R^S \approx 1/c \GeV$, which can reasonably take values in the range 0.1GeV to 10GeV, depending on $c$, with an "average" value of around $1 \GeV$. This is consistent with the range of values attainable in 
R-symmetric D-term inflation models.

                                   A fraction of the total B asymmetry, $f_{B}$, expected to be not very small compared with
 1 \cite{bbb2}, ends up in the form B-balls once the AD condensate 
collapses to a mixture of B-balls and non-relativistic squarks at $H \approx H_{i}$. The B-balls have very large charges, $B \approx 10^{23}f_{B}(1\GeV/T_{R}^{S})$, and, for $T_{R}^{S} \approx 1\GeV$, decay at a temperature $T_{d}$ in the range 1 MeV to 1 GeV \cite{bbb2}. In this case the 
B-balls will decay below the LSP freeze-out temperature, $T_{f} \approx m_{\chi}/20$, where $m_{\chi}$ is the LSP neutralino mass \cite{dmm}, and the LSP density is likely to be 
dominated by LSPs coming from B-ball decays \cite{bbb2}. 
This is particularly true in the present 
model in which there is a low reheating temperature, since, if $T_R^S < T_{f}$, the thermal relic LSP density will be strongly suppressed (by an additional factor $(T_R^S/T_f)^5$) by the period of matter domination between $T_{f}$ and $T_R^S$. 
The domination of the LSP density by B-ball decays
is, in itself, an important consequence of B-balls, requiring a
new analysis of dark matter constraints on the MSSM \cite{kedm}.
It also makes it possible to account for the 
similarity of the number densities of baryons and LSPs,
since, if $f_{B}$ is not very small compared with 1, the number density of
LSPs from B-ball decays will be naturally similar
to that of baryons \cite{bbb2}.

                       Thus, assuming nothing more than a low energy sector
 consisting of the MSSM fields together with a period of primordial D-term inflation,
 we can, in principle, naturally account for all the main observation of cosmology; the
 baryon asymmetry (via the Affleck-Dine condensate), cold dark matter 
and the baryon to dark matter number ratio (LSP neutralinos from B-ball decays) and the
homogeneity of the Universe together with the CMBR temperature fluctuations
(D-term inflation). 
This provides us with 
an economical and effective model for SUSY cosmology. 
The resulting scenario radically departs from the conventional 
radiation dominated view of SUSY cosmology at low temperatures, with the electroweak 
phase transition playing no significant role in baryogenesis and late decaying B-balls 
opening up new possibilities for SUSY dark matter. 

\subsection*{Acknowledgements}   This work has been supported by the
 Academy of Finland and by a European Union Marie Curie Fellowship under EU contract number 
ERBFM-BICT960567.

\end{document}